\documentclass[aps,floats,amssymb,twocolumn,prb, superscriptaddress]{revtex4-2}
\usepackage{calc}
\usepackage{psfrag}
\usepackage{graphicx}
\usepackage{color}
\usepackage{bm}
\usepackage{float}
\usepackage{lipsum}
\usepackage{amsmath}

\def\nn{\nonumber}
\def\bs{\boldsymbol}

\begin{document}

\title{Dipole representation of half-filled Landau level}

\author{S. Predin}
\affiliation{Scientific Computing Laboratory, Center for the Study of Complex Systems,Institute of Physics Belgrade, University of Belgrade, Pregrevica 118, 11080 Belgrade, Serbia}
\author{A. Kne{\v z}evi\'c}
\affiliation{Faculty of Physics, University of Belgrade, Studentski Trg 12-16, 11000 Belgrade, Serbia}
\author{M.V. Milovanovi\'c}
\affiliation{Scientific Computing Laboratory, Center for the Study of Complex Systems,Institute of Physics Belgrade, University of Belgrade, Pregrevica 118, 11080 Belgrade, Serbia}
\email{Corresponding author:milica.milovanovic@ipb.ac.rs}

\begin{abstract}
We introduce a variant of dipole representation for composite fermions in a half-filled Landau level, taking into account the symmetry under exchange of particles and holes. This is implemented by a special constraint on composite fermion and composite hole degree of freedom (of an enlarged space), that makes the resulting composite particle, dipole, a symmetric object. We study an effective Hamiltonian, that commutes with the constraint on the physical space, and fulfills the requirement for boost invariance on the Fermi level. The calculated  Fermi liquid parameter $F_2$ is in a good agreement with numerical investigations in 	[Phys. Rev. Lett. 121, 147601 (2018)].
\end{abstract}

\maketitle

\section{Introduction}
Fractional quantum Hall effect (FQHE) is a phenomenon of strongly correlated electrons that is amenable to quasiparticle pictures and modeling. The presence of the strong magnetic field very often leads to the dominance of the physics inside a Landau level (LL) (a subspace of the Hilbert space of the problem), and justifies approaches that assume that the description of the problem can be confined to an isolated LL. The half-filled LL problem (relevant for systems at half-filling factors) is very interesting, because it possesses an additional symmetry, not present in experiments or any other fillings of LLs, the symmetry under the exchange of particles (electrons) and holes i.e. particle-hole (PH) symmetry. On the other hand, the systems at half-filling harbor exceptional physics for FQHE: the formation of a Fermi-liquid state of composite quasiparticles (fermions) that reside in the lowest LL (LLL) \cite{hlr, wil1}, and the incompressible at the half-integer filling factor, 5/2, for electrons in the second LL \cite{wil2}. Moreover, it is widely believed that the physics at 5/2 is connected with a Cooper pairing of underlying quasiparticles that make so-called Pfaffian states \cite{mr}. The LL mixing (the influence of other LLs to the effective physics in the base LL) is very important for the physics of Pfaffian states, because it selects and stabilizes a unique Pfaffian state among three possibilities: Pfaffian, anti-Pfaffian, and PH Pfaffian. On the other hand, the LL mixing is not that important for the physics of electrons in the LLL. Moreover we are interested to know what consequences will induce the requirement for the PH symmetry that is present if we assume that the Hilbert space of the system is an isolated LL. The state in the experiments may correspond to spontaneous PH symmetry breaking state of a Hamiltonian that respects the PH symmetry. 

Even if we confine our description to the isolated LL, the problem at half-filling is still a strongly correlated one. Our hope is that by selecting, introducing appropriate quasiparticles we can, by applying approximate methods (mean-field usually), reach an effective description of the system consistent with numerical and real experiments.

The Fermi liquid (FL) concept for the physics of composite quasiparticles - composite fermions (CFs)  was introduced in the seminal work of Halperin, Lee, and Read (HLR), \cite{hlr}, on the basis of Chern-Simons field-theoretical description that does not include a projection to the LLL.  To achieve a detailed description and understanding of the Fermi-liquid state inside the LLL, Pasquier and Haldane \cite{ph}, and later Read \cite{read}, analyzed a related system of bosons at filling one, with an exact representation of the CF quasiparticles in an enlarged space of the problem. Shankar and Murthy \cite{ms} advanced, on the other hand, field-theoretical description of the dipole - CF representation of the problem \cite{ss}. In recent years a concept of a Dirac i.e. two-component quasiparticle was introduced by Son \cite{son} for the description of the half-filled LL i.e. a system that respects the PH symmetry. In Ref. \cite{goc}  a microscopic derivation of such a theory was given, where the two components were connected to the two possibilities for quasiparticles, composite fermions (CFs) and  composite holes (CHs).

Here we propose an extension of the quasiparticle view of the physics inside an isolated LL based on a one-component fermion - a dipole; a description that does not differentiate and does not distinguish CFs and CHs. We employ the enlarged-space formalism \cite{ms, ph, read}, but with a special constraint that incorporates the PH symmetry, in the system of electrons that fill a half of an isolated LL. The special constraint and demand for the boost invariance defines an effective Hamiltonian and FL description in an agreeement with numerical experiments \cite{re}.

The paper is organized as follows. In the main section, Section II, the enlarged-space formalism for the system of bosons at filling one is reviewed before the exposition of our proposal in the same section. At the end of the section we discuss the FL description based  on that proposal. In Section III, in order to further understand the nature of the introduced quasiparticles,  the problem of the bilayer i.e. two half-filled LLLs is analyzed in the new representation. In Section IV we discuss an additional implementation of our approach, a quantum Boltzmann equation based on the dipole picture that incorporates the boost invariance and PH symmetry, and end with conclusions. In Appendix A the invariance under the change of basis i.e. the $SU(N)$ invariance for the proposed formalism is described, and in  Appendix B details concerning the derivation of the quantum Boltzmann equation are given.

\section{Dipole representation in an isolated Landau level}

We study an extension of the formalism for CFs that was developed in \cite{ph,read}, for the case of bosons at filling factor one, to the case of electrons in an isolated LL that is half-filled. Although, an extension of the CF formalism to the case of half-filled LL of electrons, has to incorporate {\em bosonic} correlation holes, (i.e. artificial degrees of freedom in the set-up ) we will show that with a hard-core constraint in an enlarged space, we can delineate a physical subspace and reach a faithfull description. In the following we will review the basic formalisam for the system of bosons at filling factor one.

\subsection{Bosons at filling factor $\nu = 1$}

\subsubsection{Review of CF formalism for bosons at filling factor $\nu = 1$}

In this section we will review in short terms the CF formalism that was introduced in \cite{ph} and further developed in \cite{read} for bosons at filling factor $\nu = 1$. The CF is a composite quasiparticle of underlying system of bosons (or electrons) which consists of a boson (an electron) and its (associated) correlation hole in the incompressible or weakly compressible  (like FL state of CFs) FQHE states.  For the case of bosons at filling factor $\nu = 1$ we have one boson on average per state $\Psi_n$, in an isolated LL, $ n = 1, ... N_\phi  $. $ N_\phi $ denotes the number of flux quanta through the system. Thus $n$ enumerates states in a chosen basis of the relevant LL.

The Laughlin solution of FQHE at $\nu = 1/3$ of electrons - an excellent description of the ground state wave function, which incorporates basic correlations among particles, that are solely interaction driven in an isolated LL, motivates the introduction of the composite quasiparticles. In the regime of FQHE we can easily envision a structure of a neutral quasiparticle: an underlying boson (or an electron) and its associated correlation hole, which we can express and regard as (Laughlin) quasihole excitation (eigenstate) of the system with quantized charge and statistics (or nearly quantized and localized, almost eigenstate in a weakly compressible systems (like FL state of CFs)).
 
In the case of  bosons at filling factor $\nu = 1$ of an isolated LL, and by following the Laughlin ansatz we can easily find out that the most natural assignment for the statistics of the correlation hole is fermionic (so that the statistics of CFs is fermionic and counterbalance the Vandermonde determinant fermionic correlations, and produce overall bosonic correlations and wave function), and that it represents a deficiency of a unit of charge (a hole). Thus we are inclined to consider a composite fermionic object (CF) and associate an annihilation operator with two indecies $m$ and $n$: $c_{nm}$, where $n$ and $m$ referes to two states of a chosen basis in a LL, left index, $n$, that describes the state of an (underlying, elementary) boson and right index, $m$, that describes the state of its correlation hole - an artificial (additional) degree of freedom. In this way we are enlarging the space that we associate with the system's description; we introduce also creation operators, $ c_{m n}^{\dagger}$, so that
\begin{equation}
\{c_{n m}, c_{m' n'}^{\dagger}\}= \delta_{n,n'} \delta_{m,m'},
\end{equation}
and consider states in the enlarged space, 
\begin{equation}
c_{m n}^{\dagger} \cdots c_{p q}^\dagger \vert 0\rangle .
\label{sts}
\end{equation}
The physical subspace of this enlarged space can be delineated by projecting out artificial, non-physical degrees of freedom i.e. ``vortices" (correlation holes) that possess fermionic statistics:
\begin{equation}
\vert n_{1},\dots,n_{N}\rangle = \sum\limits_{m_{1},\dots,m_{N}}^{N_{\phi}} \varepsilon^{m_{1} \cdots m_{N}}
c_{m_{1}n_{1}}^{\dagger} \cdots c_{m_{N} n_{N}}^\dagger \vert 0\rangle
\label{perms}
\end{equation}
where $\varepsilon^{m_{1} \cdots m_{N}}$ is the Levi-Civita symbol.

We may notice that this construction is invariant under a $SU(N)$ transformation = change of basis in the LL in the R sector only, i.e.
\begin{equation}
c_{m n}^{\dagger}  \rightarrow   \sum\limits_{m'}^{N_{\phi}}  U_{m m'} c_{m' n}^{\dagger},
\end{equation}
where $ U_{m m'}$ is an $SU(N)$ matrix. (The physical states are spin-singlets under this transformation.)

In the CF representation one can consider $\rho_{m m'}^{L} $ and $\rho_{m m'}^{R}$, density operators for physical $ (L)$ and unphysical (hole-like, $R)$ degrees of freedom:
 \begin{equation}
\rho_{n n'}^{L} = \sum_{m} c_{m n}^{\dagger} c_{n' m},
\label{lgen}
\end{equation}
and
\begin{equation}
\rho_{m m'}^{R} = \sum_{n} c_{m n}^{\dagger} c_{n m'}.
\label{rgen}
\end{equation}
Also the following decomposition (of the basic state $(n, m)$ of the composite object), can be considered
\begin{equation}
c_{n m} = \int\frac{d\bs{k}}{(2 \pi)^{\frac{3}{2}}} \langle n\vert\tau_{\bs{k}}\vert m\rangle c_{\bs{k}},
\end{equation}
with $\tau_{\bs{k}} = \exp\left(i \bs{k} \cdot\bs{R}\right)$, where $\bs{R}$ is a guiding-center coordinate of a single
particle in the external magnetic field, 
\begin{equation}
[R_x , R_y ] = - i ,
\end{equation}
we took $l_B$ (magnetic length) $=1$,
and $\{\vert n\rangle\}$ are single-particle states (orbitals) in a fixed LL.

The parameter $\bs{k}$ has the meaning of the momentum of the composite object - CF. 
Physically, the state of the CF with vortex orbital $m$ and electron 
orbital $n$, can be described by a superposition of the (commutative) 
momentum ${\bs k}$ states, the weights of which depend on the effective 
distance between orbitals (the size of the dipole), $|{\bs k}_{\rm 
eff}|$, because 
 $ \tau_{\bs{k}} = \exp\left(i \bs{k} \cdot\bs{R}
\right)$ is the translation operator.

The introduced decomposition implies the following expressions for the $L$ and $R$ densities in the inverse space
\begin{equation}
\rho_{n n'}^{L} = \sum_m c_{m n}^{\dagger}c_{n' m} = \int \frac{d\bs{q}}{2\pi} \langle n'\vert \tau_{\bs{q}}\vert n\rangle
\rho_{\bs{q}}^{L},
\label{dens}
\end{equation}
where
\begin{equation}
\rho_{\bs{q}}^{L} = \int \frac{d\bs{k}}{(2\pi)^2} c_{\bs{k} - \bs{q}}^\dagger c_{\bs{k}} \exp\left(i \frac{\bs{k} \times \bs{q}}{2}\right). \label{bden}
\end{equation}
Note the inverse order of indices, $n$ and $n'$, on the left and
right hand side of (\ref{dens}). Similarly, 
\begin{equation}
\rho_{m m'}^{R} = \sum_n c_{m n}^{\dagger}c_{n m'} = \int \frac{d\bs{q}}{2\pi} \langle m\vert \tau_{\bs{q}}\vert m'\rangle
\rho_{\bs{q}}^{R},
\label{denr}
\end{equation}
where
\begin{equation}
\rho_{\bs{q}}^{R} = \int \frac{d\bs{k}}{(2\pi)^2} c_{\bs{k} - \bs{q}}^\dagger c_{\bs{k}} \exp\left(- i \frac{\bs{k} \times \bs{q}}{2}\right). \label{bdenr}
\end{equation}
The density $\rho_{\bs{q}}^{L}$ satisfies the GMP algebra,
\begin{equation}
[\rho_{\bs{q}}^{L}, \rho_{\bs{q}'}^{L}] = 2 i \sin\left(\frac{\bs{q}\times\bs{q}'}{2}\right)
 \rho_{\bs{q} + \bs{q}'}^{L},
\end{equation}
while $ \rho_{\bs{q}}^{R}$, as a density of ``holes",
\begin{equation}
[\rho_{\bs{q}}^{R}, \rho_{\bs{q}'}^{R}] = - 2 i \sin\left(\frac{\bs{q}\times\bs{q}'}{2}\right)
 \rho_{\bs{q} + \bs{q}'}^{R},
\end{equation}
i.e. GMP algebra for particles with opposite electric charge.

Thus we can realize the basic algebra of the projected to a LL electron density, as an algebra of the same density expressed via operators that represent overall neutral objects (dipoles) i.e. CFs. The CF representation, we expect will capture the basic physics of the problem, and already at mean-field level give meaningful results (stable FL if we apply Hartree-Fock (\cite{read})).

\subsubsection{Hamiltonian and constraints}

The basic Hamiltonian in the second-quantized notation,
\begin{equation}
{\cal H} = \frac{1}{2} \sum_{m_{1},\ldots, m_{4}} V_{m_{1}, m_{2}; m_{3}, m_{4}} a^\dagger_{m_1} a^\dagger_{m_2} a_{m_4} a_{m_3},
\end{equation}
can be represented  by the following Hamiltonian in CF representation \cite{read},
\begin{equation}
{\cal H} = \frac{1}{2} \sum_{\substack{m_{1},\ldots, m_{4} \\  n_{1}, n_{2}}} V_{m_{1}, m_{2}; m_{3}, m_{4}} c^\dagger_{n_1 m_1} c^\dagger_{n_2 m_2} c_{m_4 n_2} c_{m_3 n_1}.
\end{equation}
That this is possible can be seen because  by mapping bilinear $ a^\dagger_{p} a_{q} $ into $ \sum_k c^\dagger_{k p} c_{q k} $we are preserving the basic algebra of fermionic (electron) bilinears, but what may happen is that new (in the enlarged space) operators (including Hamiltonian) can map physical states into supperpositions of physical and unphysical states. Thus we have to ensure that physical states are mapped into physical (sub)space: Hamiltonian has to commute with the constraint(s) (that determine the physical subspace of the enlarged space). From (\ref{perms})  we see that in this case the cnstraint that defines the physical space is  $\rho_{n n}^R = 1 $ and $[ {\cal H}, \rho_{n n}^R ] = 0$. In the inverse space \cite{read},
\begin{eqnarray}
{\cal H} =\frac{1}{2} \int d{\bs{q}} \;{\tilde V} (|{\bs{q}}|)   \label{Ham}
: \rho^{L}({\bs{q}})
 \rho^{L}(-{\bs{q}}):,
\end{eqnarray}
 we have $[\rho_{\bs{q}}^{L}, \rho_{\bs{q}'}^{R}] = 0$, and thus $[{\cal H}, \rho_{\bs{q}}^{R}] = 0$, as required for the constraint $ \rho_{\bs{q}}^{R} = 0$. In the following section, where we study electron systems at half-fillings, the constraints will not be so simple, and we have to ensure the commutation with ${\cal H}$ at least on the physical subspace (with the help of constraints).

\subsection{Electron system of (an isolated) half-filled Landau level}

\subsubsection{Physical states}

In the case of the electron system at half-filling the correlation hole i.e. the superposition of two Laughlin quasihole constructions, if considered as an independent and well-defined degree of freedom, should carry bosonic statistics. This is a departure from the simple introduction of the unphysical degrees of freedom in the case of the bosonic system at filling one. These degrees of freedom, in that case, carry fermionic statistics and are entering the description via the simple constraint, $ \rho^R_{n n} = 1$ that ensures easy implementation of the $SU(N)$ invariance; we can transform the basis only in $R$ sector and the physical state will be invariant with respect to that particular transformation. Thus, as expected and required, if we are transforming a physical state  we have a usual, unitary implementation of $SU(N)$, and only $L$ degrees of freedom are affected. 

We may wonder if it is possible at all to implement the $SU(N)$ invariance if we have bosonic unphysical (additional) degrees of freedom that enter the CF description. If we do not consider CFs and CHs (to account for the PH symmetry in a half-filled Landau level)
and a Dirac-type theory \cite{goc}, we can attempt a description that is similar to the one in the bosonic $\nu = 1$ case, by considering only CFs, more precisely single-particle operators with two indexes, $c_{m n}$, but only of one kind, and the following associated physical states (Slater determinants) in the enlarged space: 
\begin{eqnarray}
&&\vert \Psi_{\rm phy} \rangle = \vert n_{1},\dots,n_{N_\phi/2}\rangle = \nn \\
&& \sum_{\sigma \in S_{N_\phi/2}}^{'}
c_{\sigma(m_{1}) n_{1}}^{\dagger} \cdots c_{\sigma(m_{N_\phi/2}) n_{N_\phi/2}}^\dagger \vert 0\rangle ,
\label{dets}
\end{eqnarray}
where the prime over the sum means that the sum is over permutations of distinct states, $ m_{1} \neq m_{2} \neq \cdots \neq m_{N_\phi/2} $ i.e. indexes connected with a basis in a LL: $\{ \vert n_1 \rangle, \vert n_2 \rangle, \cdots, \vert n_{N_\phi/2} \rangle, \vert m_1 \rangle, \vert m_2 \rangle, \cdots ,\vert m_{N_\phi/2} \rangle \}$.  In (\ref{dets}) we have electrons that occupy states from subspace $ V =  \{ \vert n_1 \rangle, \vert n_2 \rangle, \cdots, \vert n_{N_\phi/2}\rangle \}$ i.e. half of the available states in a LL. The second half  $ V_\bot =  \{ \vert m_1 \rangle, \vert m_2 \rangle, \cdots, \vert m_{N_\phi/2}\rangle  \}$ (states from the orthogonal subspace) are occupied by holes.

In Appendix we discuss how the $SU(N)$ invariance can be implemented   if one  consider simultaneous transformation(s) on $L$ and $R$ indecies; a transformation on only one type of index is nontrivial - nonunitary. In an isolated half-filled LL we have the particle-hole symmetry and in that case we may expect that the change of basis affects both $L$ (particle) and $R$ (hole) index, because of  the intertwined physics of particle and holes. This is unavoidable and only option we have both in the Dirac (manifestly invariant PH symmetric with CFs and CHs) or CF-only representation and description of the problem discussed here. 

\subsubsection{Hamiltonian and constraints}
The  realization of the $SU(N)$ symmetry and the quasiparticle description relies on the requirement that
\begin{equation}
\rho_{n n}^L + \rho_{n n}^R = 1, \label{con1}
\end{equation}
i.e. the hard-core constraint that we introduced at the beginning of the subsection in the description of the basic physical states (\ref{dets}). The constraint is a part of the formulation of the problem; it specifies the half-filling condition. What is special with respect to the previous introduction of the additional degrees of freedom that should represent correlation holes, here correlation holes describe hole degrees of freedom of the half-filled problem. In that sense  our approach can be considered as only effective, not microscopic, considering the way that quasiparticle description is introduced  (with respect to the bosonic case at $\nu = 1$).  But once the constraint is assumed we can build our description, i.e. effective Hamiltonian, by requiring that Hamiltonian commutes with the constraint as we will describe shortly. On the other hand, there are reasons why such constraint is appropriate. Firstly, it includes particle and hole degrees of freedom (CFs and CHs) in a way that the PH symmetry is respected. Second, it leads to a PH symmetric form of Hamiltonian, which is known as a dipole-representation of underlying quasiparticle physics introduced by Shankar and Murthy, but with an additional factor of 4 that reduces the strength of Coulomb interaction. That factor is likely the one that was missing in the interpretation of surface acoustic wave experiments by the HLR theory and the dipole based theory \cite{ss}. An additional and important reason for the form of the constraint, as we will show in the following, is that the constraint can be used to obtain a final form of the Hamiltonian that respects the boost invariance. The boost invariance should characterize an effective description at least at the Fermi level, as an invariance in the system that does not have (bare) mass (i.e. kinetic term) in its microscopic description. As we will show the calculated Fermi liquid parameter, $F_2$, on the basis of that Hamiltonian and constraint, is in a very good agreement with the numerical experiment of Ref. \cite{re}.

To get the form of Hamiltonian that respects these requirements we start with the microscopic form of Hamiltonian (the same as for bosons in (\ref{Ham})) where we abandon the requirement for normal ordering and make the following substitution,
\begin{equation}
\rho^L  (\bs{q}) \rightarrow \frac{\rho^L (\bs{q}) - \rho^R (\bs{q})}{2}. 
\end{equation}
Therefore 
\begin{eqnarray}
 H =\frac{1}{8} \int d{\bs{q}} \;{\tilde V} (|{\bs{q}}|)   \label{eHam}
 (\rho^{L}({-\bs{q}}) - \rho^{R}({-\bs{q}}))
  (\rho^{L}({\bs{q}}) - \rho^{R}({\bs{q}})). \nn \\
\end{eqnarray}
Here  $ {\tilde V} (|{\bs{q}}|) = \frac{1}{|{\bs{q}}|} \exp\left(-  \frac{|\bs{k}|^2}{2}\right) [ L_n (\frac{|\bs{k}|^2}{2}) ]^2 $, where $L_n $ represents the Laguerre polynomial for a fixed LL index $n$.
Thus we modified the Hamiltonian in the particle representation to the one that respects a PH symmetry, in the way that exchange $\rho^{L}({\bs{q}})  \leftrightarrow  \rho^{R}({\bs{q}}) $ does not change the form of the Hamiltonian. Also 
\begin{eqnarray}
[ H,  (\rho^{L}({\bs{q}}) + \rho^{R}({\bs{q}})) ]|_{ (\rho^{L}({\bs{k}}) + \rho^{R}({\bs{k}})) = 0} = 0 ,
\label{requir}
\end{eqnarray}
i.e. the constraint commutes with Hamiltonian on the physical space as required.

The Hamiltonian in (\ref{eHam}) possesses a single particle term, $ H_1 $ \cite{dose},
\begin{equation}
 H =  H_1 + :  H : ,
\end{equation}
next to the purely interacting term,  $:  H :$. If we interpret the mass in $ H_1 $ at $k_F$ as the effective mass, $ m^* $, of a FL description, we need an additional interaction, a term next to the bare one i.e. $ :  H : $ in order to achieve  (a) the FL description of the system and (b) the description that is also invariant under boosts i.e. whose Hamiltonian is purely interacting at the Fermi level. 

To implement this we may add a term that needs to represent an interaction, but at the same time to be equal to zero on the physical space (not to add or spoil energetics encoded in $ H$ based on the bare, Coulomb interaction). In the inverse space that may be term of the following form,
\begin{eqnarray}
 \int d{\bs{q}} \; C (|{\bs{q}}|)  \exp\left(-  \frac{ {\bs{q}}^2}{2}\right) [ L_n ( \frac{ {\bs{q}}^2}{2} ) ]^2 \times  \nn \\ \label{Cform}
 (\rho^{L}({-\bs{q}}) + \rho^{R}({-\bs{q}}))
  (\rho^{L}({\bs{q}}) + \rho^{R}({\bs{q}})). \nn \\
\end{eqnarray}
On the other hand, in the space of the LL orbitals, we may consider the following term,
\begin{equation}
\sum_{n, n'} \delta_{n, n'} (\rho_{n n}^L + \rho_{n n}^R  ) (\rho_{n' n'}^L + \rho_{n' n'}^R  ) , \label{2conForm}
\end{equation}
based on the constraint expressed on the space of orbitals $\{ |n \rangle \} $, which is not zero but, due to the constraint, simply a constant on the physical space, i.e. constant equal to the number of orbitals. By comparing two expressions, that constrain the form of required interaction term, we can conclude that  $C (|{\bs{q}}|) $ in (\ref{Cform}) should be a constant, independent of ${\bs{q}}$. Namely, if we regularize expression in (\ref{2conForm}) in the thermodynamic limit, by omitting the terms that do not conserve momentum in the inverse space, and may represent local single-particle potentials, we find that the remaining term, that represents an interaction invariant under translation is 
\begin{eqnarray}
 H_C = C \int d{\bs{q}} \;  \exp\left(-  \frac{ {\bs{q}}^2}{2}\right) [ L_n ( \frac{ {\bs{q}}^2}{2} ) ]^2  \times  \nn \\ \label{CFform}
 (\rho^{L}({-\bs{q}}) + \rho^{R}({-\bs{q}}))
  (\rho^{L}({\bs{q}}) + \rho^{R}({\bs{q}})), \nn \\
\end{eqnarray}
i.e. a delta-function interaction projected into an isolated LL. This term is equal to zero on the physical space.

The complete Hamiltonian that describe the low-energy physics at the Fermi level and incorporates the boost invariance is 
\begin{equation}
{\cal H}_C = H + H_C ,
\end{equation}
where the constant $C$ is chosen such that the second derivative with respect to momentum of the total single-particle dispersion in the single-particle term of ${\cal H}_C $ at the Fermi level is equal to zero. 

To probe the FL description implied by ${\cal H}_C $ we consider a generalized Coulomb interaction \cite{re},
\begin{equation}
 V _\eta (q) =  \frac{1}{|{\bs{q}}|}  \exp(- |{\bs{q}}| \eta) ,
\end{equation}
which models the effect of the finite-thickness of samples in experiments, but also stabilizes FL behavior \cite{re}. The calculated  $F_2$ - the FL parameter at angular momentum $l =2$, on the basis of the effective Hamiltonian ${\cal H}_C $, is given in Fig. 1.
\begin{figure}[H]
	\centering
	\includegraphics[scale=.18]{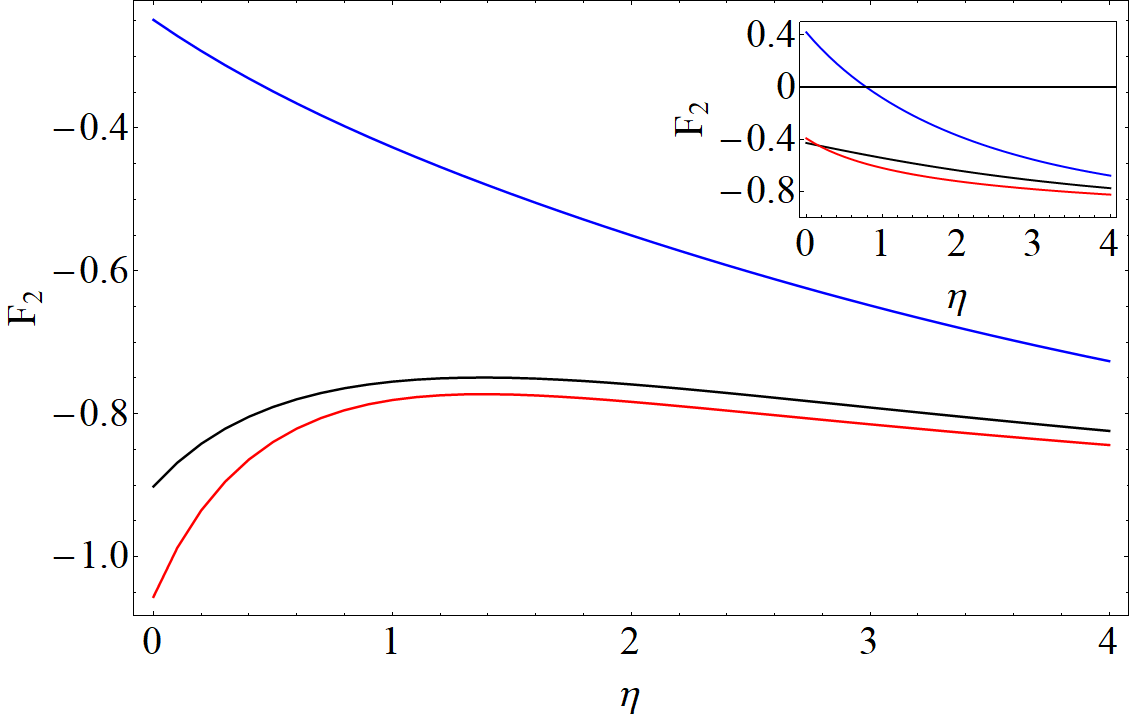}
	\caption{Fermi liquid parameter $F_2 $}
	\label{figureF2}
\end{figure}
It is in an agreement with the trends of the numerical experiment of Ref. \cite{re}; it predicts the absence and presence of the Pomeranchuk instability in the LLL and third LL respectively, for the pure Coulomb interaction, as in Ref. \cite{re}, but  does not predict one in the second LL, which exists according the analysis in \cite{re}. In the inset of Fig. 1, the values of $F_2$ are given for the usual Hamiltonian $H$ in the dipole representation, and we can see that they differ considerably from the expectations based on ${\cal H}_C $  and results of the numerical experiment in \cite{re}; the requirement for boost invariance is essential to get an agreement with results of Ref. \cite{re}. In Fig. 2 and Fig. 3 we present the values of Fermi liquid parameters, $F_3$ and $F_4$. 
\begin{figure}[H]
	\centering
	\includegraphics[scale=.18]{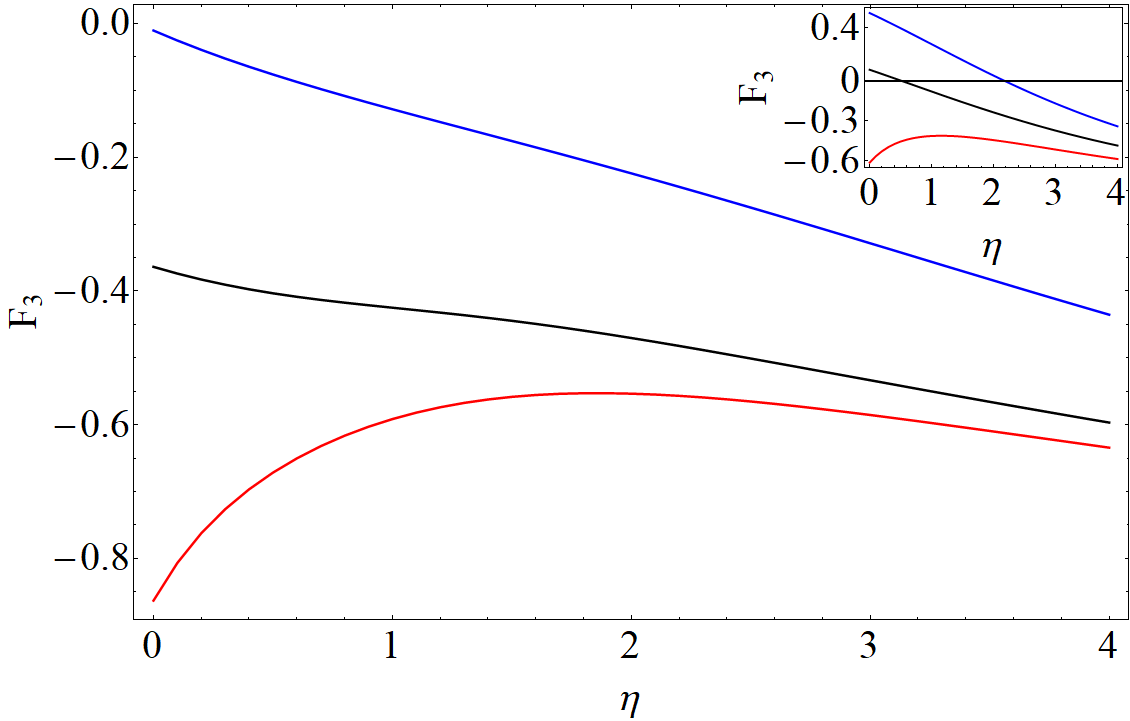}
	\caption{Fermi liquid parameter $F_3 $}
	\label{figureF3}
\end{figure}
\begin{figure}[H]
	\centering
	\includegraphics[scale=.18]{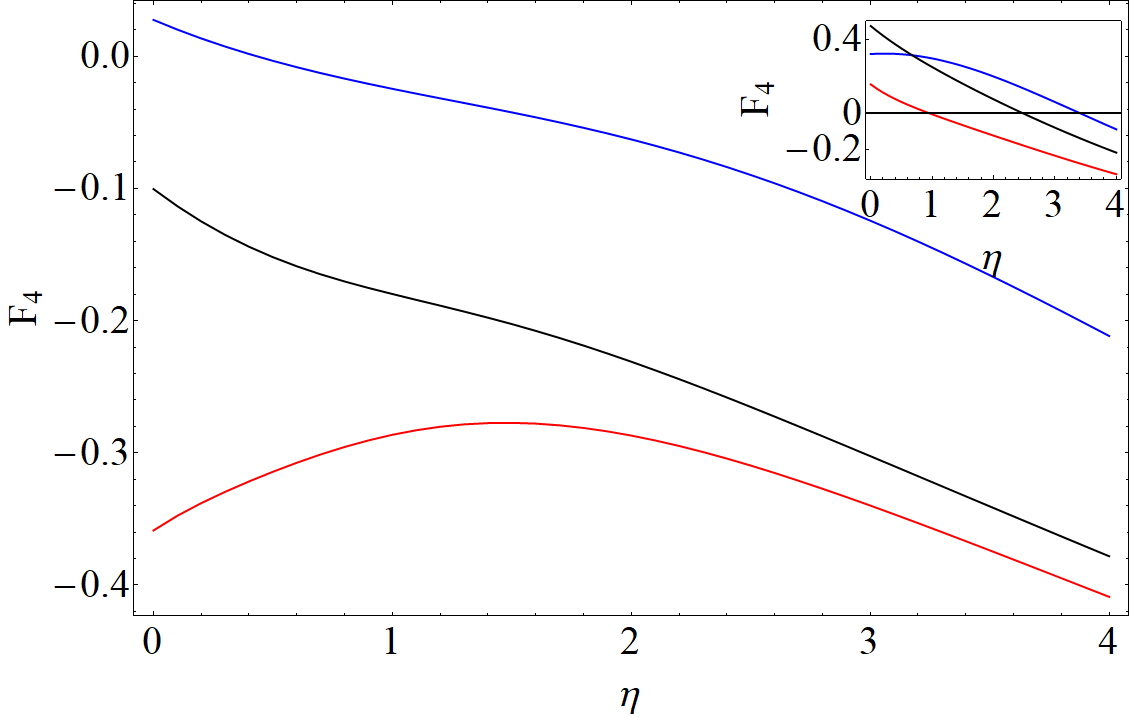}
	\caption{Fermi liquid parameter $F_4 $}
	\label{figureF4}
\end{figure}

\section{The bilayer case: pairing between two LLLs}

To get a better understanding of the underlying quasiparticle - a dipole in the new representation, we consider the quantum bilayer problem i.e. the problem with two half-filled LLLs.
The bilayer consists of two layers. Each layer represents a half-filled LLL, but with no PH symmetry, because of the interlayer interaction. The two-component physics can be represented inside a single LLL by the following Hamiltonian with electron density operators, $ \rho_{\sigma} ({\bs{q}})$, $ \sigma = \uparrow, \downarrow$ :
\begin{eqnarray}
{\cal H}_e &=& \int d{\bs{q}} \; \{ \sum_\sigma \frac{1}{2} V(|{\bs{q}}|)  \; \label{2Ham}
: \rho_{\sigma} ({\bs{q}}) \;
 \rho_\sigma (-{\bs{q}}): +  \nn \\
&& V_{\uparrow \downarrow}(|{\bs{q}}|) \;  \label{2Ham}
 \rho_{\uparrow} ({\bs{q}})\;
 \rho_\downarrow (-{\bs{q}}) \}.
\end{eqnarray}
We begin descrption of the system in the new representation by the following Hamiltonian,
\begin{eqnarray}
&& {\cal H} = \nn \\
&& \int d{\bs{q}} \; \{ \sum_\sigma \frac{1}{8} V(|{\bs{q}}|)  \; \label{22Ham}
 (\rho^{L}_\sigma ({-\bs{q}}) - \rho^{R}_\sigma ({-\bs{q}})) (\rho^{L}_\sigma ({\bs{q}}) - \rho^{R}_\sigma ({\bs{q}}))+  \nn \\
&& \frac{ V_{\uparrow \downarrow}(|{\bs{q}}|)}{4}  \;  \label{2Ham}
 (\rho^{L}_\uparrow ({-\bs{q}}) - \rho^{R}_\uparrow ({-\bs{q}})) (\rho^{L}_\downarrow ({\bs{q}}) - \rho^{R}_\downarrow ({\bs{q}}))
 \},
\end{eqnarray}
with constraints $ (\rho^{L}_\sigma ({\bs{k}}) + \rho^{R}_\sigma ({\bs{k}})) = 0; \sigma = \uparrow, \downarrow $.

We chose the form of the Hamiltonian as the one that will conform to the requirements, $[{\cal H},  (\rho^{L}_\sigma({\bs{q}}) + \rho^{R}_\sigma({\bs{q}})) ] = 0 ; \sigma = \uparrow, \downarrow $, that hold if the constraints 
 $ (\rho^{L}_\sigma({\bs{k}}) + \rho^{R}_\sigma({\bs{k}})) = 0 ; \sigma = \uparrow, \downarrow $ are applied.  The constraints define physical spaces in two layers, and also constrain the form of the Hamiltonian in the description with the enlarged space(s).
Just as in the single layer case we can add the terms that are zero on the physical space, of the form of $H_C $ in each layer to ensure the boost invariance of the system at the Fermi level. Our main interest is the effect of  the interlayer interaction. To get a transparent representation of the underlying physics, within the physical space, we can add effectively  zero terms, and transform the operators that define the interlayer interaction in the following way,
\begin{eqnarray}
&&  (\rho^{L}_\uparrow ({-\bs{q}}) - \rho^{R}_\uparrow ({-\bs{q}})) (\rho^{L}_\downarrow ({\bs{q}}) - \rho^{R}_\downarrow ({\bs{q}})) \nn \\
&& -  (\rho^{L}_\uparrow ({-\bs{q}}) + \rho^{R}_\uparrow ({-\bs{q}})) (\rho^{L}_\downarrow ({\bs{q}}) + \rho^{R}_\downarrow ({\bs{q}}))  \nn \\
&& =  - 2 [\rho^{L}_{\uparrow} ({-\bs{q}})\;
 \rho^{R}_\downarrow ({\bs{q}}) +
\rho^{R}_{\uparrow} ({-\bs{q}})\;
 \rho^{L}_\downarrow ({\bs{q}})] . \label{rep}
\end{eqnarray}
Now the effective form of the interlayer interaction represents a view of the underlying physics: excitonic binding of electrons and holes, i.e. CFs and CHs as emphasized in the recent work in Ref. \cite{hs}, which we know is a completely justified view of the physics at small distances. But  here we have only one effectively neutral quasiparticle operator ``c"  which is CF and CH at the same time - a simple dipole; the excitonic pairing that is implied by (\ref{rep})  is effectively Cooper pairing of the underlying quasiparticles ``c"  from each layer. Indeed, in the BCS mean-field treatment, we have an obvious instability described by the order parameter $\Delta_{\bs{k}}$,
\begin{equation}
\Delta_{\bs k} =  \int \frac{d\bs{q}}{(2\pi)^2}  V_{\uparrow \downarrow}(| {\bs q} - {\bs k}| )   \frac{\Delta_{\bs q}}{2 E_{\bs q}}, \label{bilayerbcs}
\end{equation}
where
\begin{equation}
V_{\uparrow \downarrow}(q) = \frac{\exp\left( - q d \right)}{q} \exp\left( - \frac{q^2}{2} \right),
\end{equation}
and  $E_{{\bs q}}$ is the Bogoliubov quasiparticle energy, defined considering the complete Hamiltonian that respects the boost invariance, and $d$ is the distance between the layers. 
\begin{figure}[H]
	\centering
	\includegraphics[scale=.5]{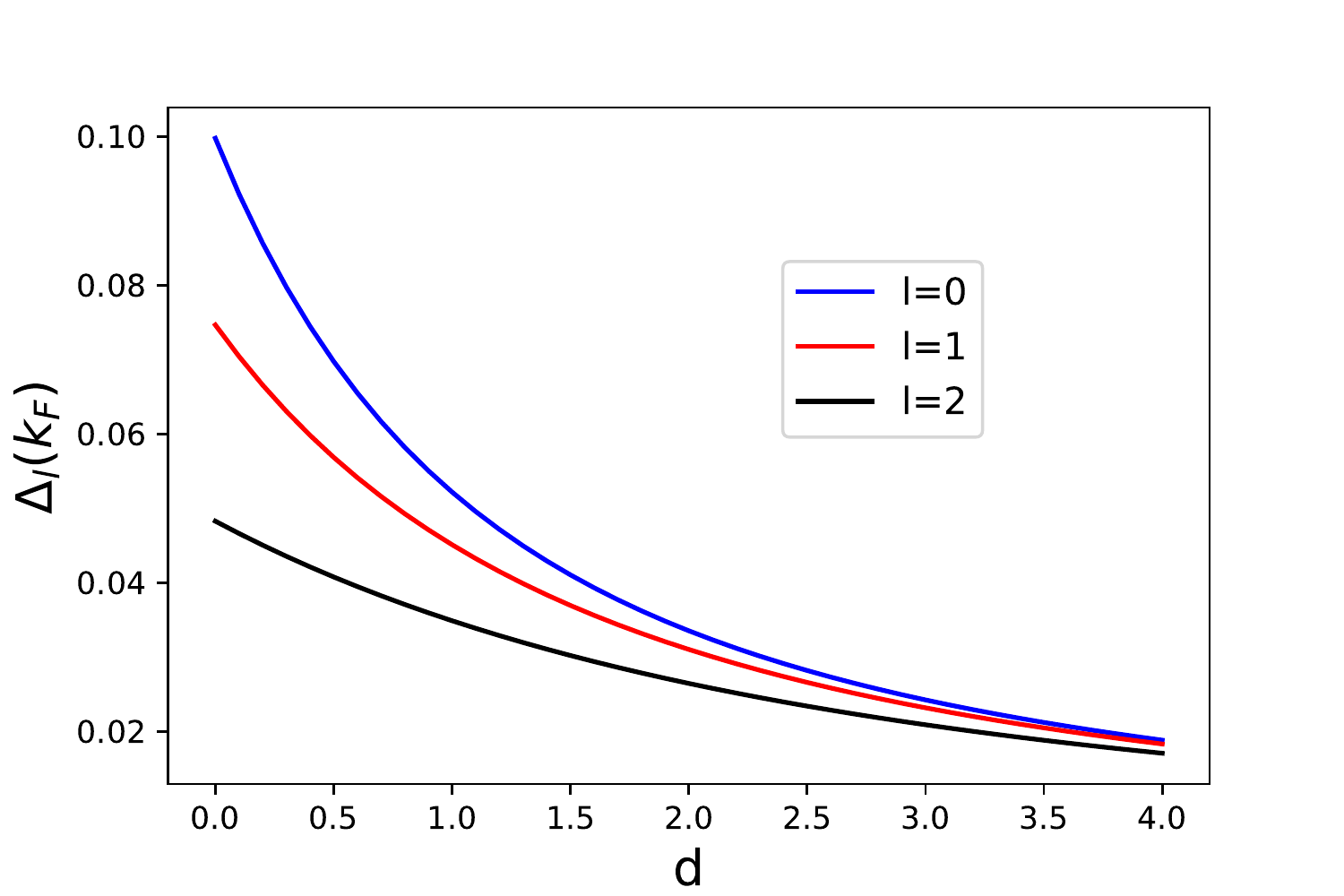}
	\caption{The solutions for $\Delta_{\bs k} $ calculated selfconsistently  using (\ref{bilayerbcs}), in the case of $s $, $p $, and $d $ wave at $ k = k_F$. }
	\label{figure4}
\end{figure}

In Fig. 4, solutions for $\Delta_{\bs k} $ at $ k = k_F$ for $s $, $p $, and $d $ wave are plotted as a function of distance. For a large interval of $d$, the results are in a qualitative agreement with the most recent numerical results of Ref. \cite{hs}, done on the sphere, when we identify CF - CH (excitonic) pairing of the same reference with the $s$-wave Cooper pairing of dipoles in the representation that we presented here. Nevertheless, a question can be raised whether for larger (or even intermediate, $ d \sim l_B ) $ distances, an effective description given by the Hamiltonian in (\ref{22Ham}) is more appropriate, because at very large distances we expect two decoupled Fermi liquids, for which dipole representation, with dipole densities, $ \rho^{L}_\sigma ({\bs{q}}) - \rho^{R}_\sigma ({\bs{q}}), $
$\sigma = \uparrow, \downarrow $ that are interacting, seems quite appropriate (according to our reasoning and comments in Section II B). Then the mean field solutions $\Delta_{\bs k}^s $, for dominant  $s$ wave excitonic (or $p$ wave Cooper ) instability, for this set up, will necessarily behave near $k = 0$ as  $\Delta_{\bs k}^s \sim |{\bs k}|^2 \;\; (|\Delta_{\bs k}^p | \sim |{\bs k}|^3 ) $, and thus exemplify an anomalous behavior and may lead to the prediction of a new (intermediate) phase detected in numerical experiments on torus \cite{fd,mp}.



\section{Discussion and Conclusions}

To further gauge the FL nature of our system (in the proposed formalism) we can consider the quantum Boltzmann equation for the Wigner function,
\begin{equation}
\nu ({\bs k}, {\bs r}) = \int d\bs{s} \exp\left( i {\bs k}{\bs s} \right)   {\rm Tr}\{ \rho \Psi^\dagger ({\bs r}  + \frac{{\bs s}}{2} ) \Psi ({\bs r}  - \frac{{\bs s}}{2} ) \},
\end{equation}
where $\rho$ is the density matrix of the system, and $ \Psi , \Psi^\dagger $ are second-quantized operators that are defined, in the long-distance approximation (in the usual way) on the space of
commuting coordinates,
\begin{eqnarray}
&& \Psi ({\bs x}) = \int \frac{d\bs{k}}{(2\pi)^2} \exp\left( i {\bs k}{\bs x} \right)  c_{\bs k},  \nn \\
&& \Psi^\dagger ({\bs x}) = \int \frac{d\bs{k}}{(2\pi)^2} \exp\left( - i {\bs k}{\bs x} \right)  c^\dagger_{\bs k}.
\end{eqnarray}
Applying $ i  \frac{\partial \rho}{\partial t} = [ {\cal H}_C , \rho ] $, i.e. von Neumann equation, we arrive at the following equation for  $\nu ({\bs k}, {\bs r}) $,
\begin{equation}
i  \frac{\partial \nu ({\bs k}, {\bs r})}{\partial t}  =\int d\bs{s}\exp\left( i {\bs k}{\bs s} \right)   {\rm Tr}\{ \rho [ \Psi^\dagger ({\bs r}  + \frac{{\bs s}}{2} ) \Psi ({\bs r}  - \frac{{\bs s}}{2} ) ,{\cal H}_C ]  \}. \label{eqforqbe}
\end{equation}
By considering shifts in ${\bs r}$ in single-particle correlators  to linear order, we can derive an effective expression presented in Appendix B. Only in the small-momentum limit i.e. the limit in which change of momentum (that couples with the ${\bs r}$ coordinate) is small, the expression takes the form of the usual quantum Boltzmann equation for a description of a FL with a boost invariance,
\begin{eqnarray}
&& \frac{\partial \nu}{\partial t} + \partial_{\bs r} \nu   \partial_{\bs k} \epsilon - \partial_{\bs k} \nu   \partial_{\bs r} \epsilon = {\rm additional} \nn \\
&&  {\rm \;\;\;\;\;\; (negligable\; in\; small\; momentum\; transfer )\; terms}. \nn \\ \label{qbe}
\end{eqnarray}
Here
\begin{eqnarray}
\epsilon =
-  \int  d{\bs q} \exp\left(  - \frac{{\bs q}^2}{2}\right)
 ( \frac{V({\bs q})}{4} - 2 C ) ({\bs q} \times {\bs k})^2 \nu ({\bs k} + {\bs q}), \nn \\
\end{eqnarray}
i.e. the Fock contribution to the quasiparticle dispersion that also includes the $C$-interaction contribution i.e. the interaction defined in (\ref{CFform}). The $C$-interaction ensured that no term with a finite mass $(= \frac{1}{M} {\bs k}  \nabla_{\bs r}  \nu )$ appears in the quantum Boltzmann equation and on the Fermi level to the order that was considered. Overall the description based on the dipole representation is in accordance with Ref. \cite{klw} and the quantum Boltzmann equation for CFs in the absence of the projection to a LL (based on the Chern-Simons field-theoretical approach), if we associate the processes behind smooth variations of Fermi surface with small-momentum transfer ones in the dipole representation that lead to a FL behavior. We leave  a detailed comparison and analysis of the quantum Boltzmann equation for future work. 

Thus the $C$-interaction ensures the boost invariance even if we go beyond the ordinary (Hartree)-Fock mean-field approach. Our description incorporates also the PH symmetry, and does not make a potential bias to CFs like the use of the Rezayi-Read state \cite{rere}  in
\cite{re}. Thus the discrepancy between our prediction for  the absence of the Pomeranchuk instability in the second LL with respect to Ref. \cite{re} may come from the explicit PH symmetry breaking in their analysis. The PH symmetry breaking can be associated with LL mixing, which may drive the Pomeranchuk instability and  also stabilize $p$-wave at larger distances in the bilayer system.

Our proposal keeps the PH symmetry, by the application of the constraint in (\ref{con1}). At the same time it places correlation hole in the place of real holes, which seems an unusual circumstance - correlation holes should bind to particles (or vice versa). But this is a reflection of the effective physics at the Fermi level where particles are shifted from their correlation holes by the amount proportional to $ k_F l_B^2 = l_B $, i.e.  on the order of average distance between LL orbitals. This is an important consequence of the projection to a fixed LL of the Fermi sea correlations among quasiparticles \cite{rere}. 

Therefore our proposal takes into account the requirements for the boost invariance and PH symmetry in the theory that captures the effective physics at the Fermi level  of dipole quasiparticles.
\section{Acknowledgments} We thank Antun Bala{\v z} for discussions on related problems. We acknowledge useful discussions with Jak{\v s}a Vu{\v c}i{\v c}evi{\' c}. Computations
were performed on the PARADOX supercomputing facility (Scientific Computing Laboratory, Center for the Study of Complex Systems, Institute of Physics Belgrade). S. P. and M. V. M. acknowledge funding provided by the Institute of Physics Belgrade, through the grant by the Ministry of Science, Technological Development, and Innovations of the Republic of Serbia. S.P. acknowledges funding supported as returning expert by the Deutsche Gesellschaft für Internationale Zusammenarbeit (GIZ) on behalf of the German Federal Ministry for Economic Cooperation and Development (BMZ).

\appendix

\section{The $SU(N)$ invariance in the half-filled case}
We may ask ``How the $SU(N)$ invariance can be implemented on these states?" First let's consider an artificial but instructive problem of electrons at $\nu = 1$ in a CF representation. The unique physical state can be defined in an enlarged space by a trace on bosonic - artificial degrees of freedom,
\begin{eqnarray}
\vert \Psi_{\rm phy} \rangle^{\nu =1}  = \sum_{\sigma \in S_{N_\phi}}^{'}
c_{\sigma(m_{1}) n_{1}}^{\dagger} \cdots c_{\sigma(m_{N_\phi}) n_{N_\phi}}^\dagger \vert 0\rangle ,
\label{dets2}
\end{eqnarray}
and $ m_{1} \neq m_{2} \neq \cdots \neq m_{N_\phi} $. Thus bosonic (artificial) degrees of freedom enter as  hard-core bosons into the description. The requirement is a necessary condition for the implementation of the $SU(N)$ invariance.  Namely, by introducing
\begin{equation}
c_{m n}^{\dagger} \rightarrow  \sum\limits_{m'}^{N_{\phi}}  U_{m m'} c_{m' n}^{\dagger} ,
\label{sutrans}
\end{equation}
we may notice that under the hard-core constraint the $SU(N)$ transformation in the $R$ sector will act locally - it will induce a number - the permanent of the $SU(N)$ matrix that will multiply the same state:
\begin{equation}
{\hat g}^{R} \vert \Psi_{\rm phy} \rangle^{\nu =1} = \Lambda (g) \vert \Psi_{\rm phy} \rangle^{\nu =1}.
\end{equation}
The $SU(N)$ invariance exists if its action is unitary under simultaneous transformations in $R$ and $L$ sectors. Thus allowing a nonunitary (in general) action on $L$ degrees of freedom,
\begin{equation}
{\hat g}^{L} \vert \Psi_{\rm phy} \rangle^{\nu =1} = \frac{1}{ \Lambda (g)}  \vert \Psi_{\rm phy} \rangle^{\nu =1},
\end{equation}
we can reach the invariance:
\begin{equation}
 {\hat g}_{SU(N)} \vert \Psi_{\rm phy} \rangle^{\nu =1}= {\hat g}^{L} {\hat g}^{R} \vert \Psi_{\rm phy} \rangle^{\nu =1} = \vert \Psi_{\rm phy} \rangle^{\nu =1}.
\end{equation}

The previous case is artificial and of no physical importance but suggests how the $SU(N)$ invariance can be accommodated in the system of interest: half-filled LL of electrons. We may begin with the usual transformation on $R$ indexes as in (\ref{sutrans}). As a result of the hard-core constraint among holes we have:
\begin{eqnarray}
&&\vert \Psi_{\rm phy} \rangle^R =  \nn \\
&& \sum_{ \{m_{1}' \neq m_{2}' \neq \cdots \neq m_{N_\phi/2}' \}} \sum_{\sigma \in S_{N_\phi/2}}   \nn \\
&&[ \sum_{p \in S_{N_\phi/2}}^{'} U^{p(m_1) \sigma(m_1^{'})} \cdots U^{p (m_{N_\phi/2})  \sigma(m_{N_\phi/2}^{'})}] \nn \\
&& c_{\sigma(m_{1}^{'}) n_{1}}^{\dagger} \cdots c_{\sigma(m_{N_\phi/2}^{'}) n_{N_\phi/2}}^\dagger \vert 0\rangle ,
\end{eqnarray}
where the first sum is over all possible distinct collections  of  $N_\phi/2$ numbers i.e. basis vectors. The number in the square brackets we can denote by
$ [\cdots] = \Lambda ( \{m_i^{'} \}) = K (\{ n_i^{'} \})$ where $\{n_i^{'} \}$ denote basis states from the subspace orthogonal to the one spanned by $\{m_i^{'} \}$.  
The number $\Lambda ( \{m_i^{'} \})$ is symmetric under permutations of $\{m_i^{'} \}$ and can be pulled out of the sum over $\sigma$ permutations. Thus
\begin{eqnarray}
&&{\hat g}^{R} \vert \Psi_{\rm phy} \rangle \equiv \vert \Psi_{\rm phy} \rangle^R \nn \\
&& = \sum_{ \{m_{1}' \neq m_{2}' \neq \cdots \neq m_{N_\phi/2}' \}} \Lambda ( \{m_i^{'} \})  \nn \\
&& \sum_{\sigma \in S_{N_\phi/2}}   c_{\sigma(m_{1}^{'}) n_{1}}^{\dagger} \cdots c_{\sigma(m_{N_\phi/2}^{'}) n_{N_\phi/2}}^\dagger \vert 0\rangle
\end{eqnarray}
Now if we define a $SU(N)$ transformation on the $L$ indecies, in such a way that for each term  $ \{n_i \}  \rightarrow  \{n_i^{'} \}$ in the generated expansion we divide by $K (\{ n_i^{'} \})$ i.e.
\begin{eqnarray}
&&{\hat g}^{L} \vert \Psi_{\rm phy} \rangle \equiv \vert \Psi_{\rm phy} \rangle^L \nn \\
&& = \sum_{ \{n_{1}' \neq n_{2}' \neq \cdots \neq n_{N_\phi/2}' \}}  \frac{1}{K (\{ n_i^{'} \})} \nn \\
&&  U^{n_1^{'} n_1 } \cdots U^{ n_{N_\phi/2}^{'} n_{N_\phi/2} } \nn \\
&& \sum_{\sigma \in S_{N_\phi/2}}   c_{\sigma(m_{1}) n_{1}^{'}}^{\dagger} \cdots c_{\sigma(m_{N_\phi/2}) n_{N_\phi/2}^{'}}^\dagger \vert 0\rangle,
\end{eqnarray}
it follows
\begin{eqnarray}
&& {\hat g}_{SU(N)} \vert \Psi_{\rm phy} \rangle= {\hat g}^{L} {\hat g}^{R} \vert \Psi_{\rm phy} \rangle = \nn \\
&& = \sum_{ \{n_{1}' \neq n_{2}' \neq \cdots \neq n_{N_\phi/2}' \}}   \nn \\
&&  U^{n_1^{'} n_1 } \cdots U^{ n_{N_\phi/2}^{'} n_{N_\phi/2} } \nn \\
&& \sum_{\sigma \in S_{N_\phi/2}}   c_{\sigma(m_{1}^{'}) n_{1}^{'}}^{\dagger} \cdots c_{\sigma(m_{N_\phi/2}^{'}) n_{N_\phi/2}^{'}}^\dagger \vert 0\rangle
\label{impl}
\end{eqnarray}
under application of all hard-core constraints that define the physical space in the enlarged space. Any change of basis in enlarged space (that acts on $R$ and $L$ indexes) is represented  on physical states by the unitary implementation (\ref{impl}) as required and expected. The implementation is unitary and represents an expected expansion on physical states, but cannot be described as a simple action on $L$ indexes.

\section{Quantum Boltzmann equation}

The explicit expression for the r.h.s. of Eq. (\ref{eqforqbe}) to the linear order in shifts in  single-particle correlators is
\begin{eqnarray}
i  \frac{\partial \nu ({\bs k}, {\bs r})}{\partial t}  =\;\;\;\;\;\;\;\;\;\;\;\;\;\;\;\;\;\;\;\;\;\;\;\;\;\;\;\;\;\;\;\;\;\;\;\;& \nn \\
\int \frac{d\bs{q}}{(2\pi)^2}  \frac{{\tilde V}({\bs q})}{4} \{ ({\bs q} \times {\bs k}) (i {\bs q} \times \nabla_{\bs r})  [ \nu({\bs k}, {\bs r} , t ) \nu({\bs k} + {\bs q}, {\bs r} , t ) ]& \nn \\
- \frac{1}{4} [i \nabla_{\bs k} (i {\bs q} \times \nabla_{\bs r}) \nu({\bs k}, {\bs r} , t )]  [ \nabla_{\bs r} (i {\bs q} \times \nabla_{\bs r}) \nu({\bs k} + {\bs q}, {\bs r} , t )]  \nn \\
+ \frac{1}{4} [ \nabla_{\bs r} (i {\bs q} \times \nabla_{\bs r}) \nu({\bs k}, {\bs r} , t )]  [i \nabla_{\bs k} (i {\bs q} \times \nabla_{\bs r}) \nu({\bs k} + {\bs q}, {\bs r} , t )]  \nn \\
+ i ({\bs q} \times {\bs k})^2 [ \nabla_{\bs r} \nu({\bs k}, {\bs r} , t ) \nabla_{\bs k} \nu({\bs k} + {\bs q}, {\bs r} , t ) -  \;\;\;\;\;\;\;\;\;\;\;\;\;\;\;\; \;\;\;\nn \\
\;\;\;\;\;\;\;\;\;\;\;\nabla_{\bs k} \nu({\bs k}, {\bs r} , t ) \nabla_{\bs r} \nu({\bs k} + {\bs q}, {\bs r} , t ) ] \nn \\
+  ({\bs q} \times {\bs k}) (i {\bs q} \times {\bs z})  [ \nu({\bs k}, {\bs r} , t ) \; \nabla_{\bs r} \nu({\bs k} + {\bs q}, {\bs r} , t ) - \;\;\;\;\;\;\;\;\;\;\;\;\; \nn \\
\;\;\;\;\;\;\;\;\;\;\;\nabla_{\bs r} \nu({\bs k}, {\bs r} , t )  \;  \nu({\bs k} + {\bs q}, {\bs r} , t ) ] \} \nn \\
\int \frac{d\bs{q}}{(2\pi)^2}  (-2C) \{ ({\bs q} \times {\bs k}) (i {\bs q} \times \nabla_{\bs r})  [ \nu({\bs k}, {\bs r} , t ) \nu({\bs k} + {\bs q}, {\bs r} , t ) ] \nn \\
+  i ({\bs q} \times {\bs k})^2 [ \nabla_{\bs r} \nu({\bs k}, {\bs r} , t ) \nabla_{\bs k} \nu({\bs k} + {\bs q}, {\bs r} , t ) - \;\;\;\;\;\;\;\;\;\;\;\;\;\;\; \nn \\
\;\;\;\;\;\;\;\;\;\;\;\nabla_{\bs k} \nu({\bs k}, {\bs r} , t ) \nabla_{\bs r} \nu({\bs k} + {\bs q}, {\bs r} , t ) ] \nn \\
- \frac{i}{8} [ \nabla_{\bs k}  \nu({\bs k}, {\bs r} , t )  \nabla_{\bs r} ({\bs q} \times \nabla_{\bs r})^2 \nu({\bs k} + {\bs q}, {\bs r} , t ) \;\;\;\;\;\;\;\;\;\;\;\;\;\;\;\; \nn \\
+ \nabla_{\bs k} ({\bs q} \times \nabla_{\bs r})^2 \nu({\bs k}, {\bs r} , t )  \nabla_{\bs r}  \nu({\bs k} + {\bs q}, {\bs r} , t ) ]  \nn \\
+ \frac{i}{8} [ \nabla_{\bs r}  \nu({\bs k}, {\bs r} , t )  \nabla_{\bs k} ({\bs q} \times \nabla_{\bs r})^2 \nu({\bs k} + {\bs q}, {\bs r} , t ) \;\;\;\;\;\;\;\;\;\;\;\;\;\;\;\; \nn \\
+ \nabla_{\bs r} ({\bs q} \times \nabla_{\bs r})^2 \nu({\bs k}, {\bs r} , t )  \nabla_{\bs k}  \nu({\bs k} + {\bs q}, {\bs r} , t ) ]  \nn \\
+ \frac{i}{4} [ \nu({\bs k} + {\bs q}, {\bs r} , t )  ({\bs q} \times {\bs k})  ({\bs q} \times \nabla_{\bs r}) \nu({\bs k}, {\bs r} , t ) - \;\;\;\;\;\;\;\;\;\;\; \nn \\
 \nu({\bs k}, {\bs r} , t )  ({\bs q} \times {\bs k})  ({\bs q} \times \nabla_{\bs r}) \nu({\bs k} + {\bs q} , {\bs r} , t ) ] \}. \nn \\
\end{eqnarray}
If we neglect $({\bs q} \times \nabla_{\bs r})$ w.r.t. $({\bs q} \times {\bs k}) $ and higher in $ \nabla_{\bs r}$ contributions, and also calculate the overall contribution to the mass term ($ \sim {\bs k} \nabla_{\bs r}$ ), in the small $ {\bs q} $ limit,  we come to the simple form in (\ref{qbe}).



\end{document}